\documentstyle[12pt,epsf]{article}
\title{Magnetic properties
       \protect\\
       of periodic nonuniform spin-$\frac{1}{2}$ $XX$ chains
       \protect\\
       in a random Lorentzian transverse field}
\author{Oleg Derzhko\\
\small {\em {Institute for Condensed Matter Physics,}}\\
\small {\em {1 Svientsitskii St., L'viv-11, 290011, Ukraine}}}
\date{\today}

\begin{document}

\maketitle

\begin{abstract}
Using continued fractions
we examine the density of states,
transverse magnetization
and static transverse linear susceptibility
of a few periodic nonuniform spin-$\frac{1}{2}$ $XX$ chains
in a random Lorentzian transverse field.
\end{abstract}

\vspace{1cm}


\vspace{1cm}


\vspace{1mm}

\noindent
{\bf Postal address:}\\
{\em
Dr. Oleg Derzhko\\
Institute for Condensed Matter Physics\\
1 Svientsitskii St., L'viv-11, 290011, Ukraine\\
Tel: (0322) 42 74 39\\
Fax: (0322) 76 19 78\\
E-mail: derzhko@icmp.lviv.ua

\clearpage

\renewcommand\baselinestretch {1.35}
\large\normalsize

The one-dimensional spin-$\frac{1}{2}$ $XY$ model
was introduced in 1961 by Lieb, Schultz and Mattis.$^{1}$
The authors of that celebrated paper found
that several statistical mechanics calculations
for that spin model could be performed exactly
because
it can be rewritten
as a model of noninteracting spinless fermions with the
help of the Jordan-Wigner transformation.
Evidently, the formulation of the spin-$\frac{1}{2}$ $XY$ chain
in terms of fermions permits one to give a magnetic interpretation
to the results
derived for one-dimensional tight-binding spinless fermions.
As an example of where this
relationship has been exploited
one can cite the papers
on the Lloyd model$^{2,3}$ and corresponding papers on
the spin-$\frac{1}{2}$ $XX$ chain
in a random Lorentzian transverse field.$^{4,5}$
The work reported in the present paper has been inspired by some
results on one-dimensional
tight-binding Hamiltonians
for periodically modulated lattices$^{6,7}$
and
spinless Falicov-Kimball
model.$^{8,9}$
Combining the approach developed in those papers
and the treatment of the Lloyd model presented in Refs. 2, 3
we shall calculate exactly
the random-averaged one-fermion Green functions
(that yield the density of states and therefore the thermodynamics)
for the periodic nonuniform
spin-$\frac{1}{2}$ $XX$ chain
in a random Lorentzian transverse field.
We shall treat a few particular
chains in order to discuss the changes in the density of states
and magnetic properties
induced by periodic nonuniformity and diagonal disorder.
It should be noted, in passing,  that the periodic nonuniform
spin-$\frac{1}{2}$ $XX$ chain was considered
in several papers
dealing with the spin-Peierls instability
in a spin-$\frac{1}{2}$ $XX$ chain$^{10,11}$
(see also Refs. 12-16).
However, those papers concentrated
on the influence of the structural degrees of freedom
upon the magnetic ones,
rather than on the properties of a magnetic chain
with regularly alternating exchange couplings.
One should also mention
a study of a spin-$\frac{1}{2}$ $XX$ model
on a one-dimensional superlattice$^{17}$
(such a model can be viewed as
a nonuniform chain with periodically varying
exchange coupling)
but consideration was restricted to the excitation spectrum.
Our communication is also related to the work in Ref. 18
and may be viewed as a further study
of the effects of periodic nonuniformity and randomness on
magnetic properties of spin-$\frac{1}{2}$ chains.

Let us consider a cyclic
nonuniform $XX$ chain of $N\to\infty$ spins $\frac{1}{2}$
in a transverse field described by the Hamiltonian
\begin{eqnarray}
H=\sum_{n=1}^{N}\Omega_ns_n^z
+2\sum_{n=1}^{N}I_n\left(s^x_ns^x_{n+1}+s^y_ns^y_{n+1}\right)
\nonumber\\
=\sum_{n=1}^{N}\Omega_n\left(s_n^+s_n^--\frac{1}{2}\right)
+\sum_{n=1}^{N}I_n\left(s^+_ns^-_{n+1}+s^-_ns^+_{n+1}\right).
\end{eqnarray}
Here $\Omega_n$ is the transverse field at the site $n$, and
is assumed to be a random variable with the Lorentzian
probability distribution
\begin{eqnarray}
p(\Omega_n)=\frac{1}{\pi}
\frac{\Gamma_n}{(\Omega_n-\Omega_{0n})^2+\Gamma_n^2},
\end{eqnarray}
$\Omega_{0n}$ is the mean value of the transverse field
at the site $n$,
and $\Gamma_n$ is the width of its distribution.
$2I_n$ is the exchange coupling between the sites $n$ and $n+1$.
After making use of the Jordan-Wigner transformation
the model is recasted into a chain of spinless fermions
governed by the Hamiltonian
\begin{eqnarray}
H=\sum_{n=1}^{N}\Omega_n\left(c_n^+c_n-\frac{1}{2}\right)
+\sum_{n=1}^{N}I_n\left(c^+_nc_{n+1}-c_nc^+_{n+1}\right)
\end{eqnarray}
(the boundary term that is non-essential for the calculation
of thermodynamic quantities
has been omitted).
Note, that for the non-random case ($\Gamma_n=0$) assuming the transverse
field in (3) to be uniform one arises at the Hamiltonian
considered in Ref. 6.
In addition, after substitution
$\Omega_n\rightarrow Uw_n,$
$I_n\rightarrow -t$,
Eq. (3) transforms into the
Hamiltonian of a one-dimensional spinless Falicov-Kimball model
in the notations used in Refs. 8, 9.
In model (3), considered here,
not only the transverse fields
that are independent random Lorentzian variables
having the mean values and widths of their distribution, but also
the exchange couplings between neighbouring spins,
vary from site to site.

Let us introduce the temperature double-time Green functions
\linebreak
$G_{nm}^{\mp}(t)
=\mp{\mbox{i}}\theta(\pm t)
\langle\left\{c_n(t), c_m^+(0)\right\}\rangle,$
$G_{nm}^{\mp}(t)=\left(1/2\pi\right)
\int_{-\infty}^{\infty}{\mbox{d}}\omega
\exp\left(-{\mbox{i}}\omega t\right)
G_{nm}^{\mp}(\omega\pm{\mbox{i}}\epsilon),$
$\epsilon\rightarrow+0,$
where the angular brackets denote the thermodynamic average.
Utilising a set of equations for
$G_{nm}^{\mp}\equiv G_{nm}^{\mp}(\omega\pm{\mbox{i}}\epsilon)$,
and performing random averaging using contour integrals$^{2-5}$,
one finds the following set of equations for the random-averaged Green
functions
\begin{eqnarray}
\left(\omega\pm{\mbox{i}}\Gamma_n-\Omega_{0n}\right)
\overline{G_{nm}^{\mp}}
-I_{n-1}\overline{G_{n-1,m}^{\mp}}
-I_{n}\overline{G_{n+1,m}^{\mp}}
=\delta_{nm}.
\end{eqnarray}
Here $\overline{(\ldots)}\equiv
\int_{-\infty}^{\infty}{\mbox{d}}\Omega_1p(\Omega_1)
\ldots
\int_{-\infty}^{\infty}{\mbox{d}}\Omega_Np(\Omega_N)
(\ldots).$
Our task is to evaluate the diagonal random-averaged Green functions
$\overline{G_{nn}^{\mp}}$,
the imaginary parts
of which give the random-averaged density of states
$\overline{\rho(\omega)}$,
\begin{eqnarray}
\overline{\rho(\omega)}
=\mp\frac{1}{\pi N}
\sum_{n=1}^N{\mbox{Im}}\overline{G_{nn}^{\mp}}
\end{eqnarray}
that in its turn, yields thermodynamic properties of
spin model (1).
It is a simple matter to obtain
from Eq. (4) the following representation for
$\overline{G_{nn}^{\mp}}$
\begin{eqnarray}
\overline{G_{nn}^{\mp}}
=\frac{1}
{\omega\pm{\mbox{i}}\Gamma_n-\Omega_{0n}-\Delta^-_n-\Delta^+_n},
\nonumber\\
\Delta^-_n=\frac{I_{n-1}^2}
{\omega\pm{\mbox{i}}\Gamma_{n-1}-\Omega_{0,n-1}-
\frac{I_{n-2}^2}
{\omega\pm{\mbox{i}}\Gamma_{n-2}-\Omega_{0,n-2}-_{\ddots}}},
\nonumber\\
\Delta^+_n=\frac{I_{n}^2}
{\omega\pm{\mbox{i}}\Gamma_{n+1}-\Omega_{0,n+1}-
\frac{I_{n+1}^2}
{\omega\pm{\mbox{i}}\Gamma_{n+2}-\Omega_{0,n+2}-_{\ddots}}}.
\end{eqnarray}
Eqs. (5), (6) are extremely useful
for examining thermodynamic properties
of periodic nonuniform
spin-$\frac{1}{2}$ $XX$ chain in a random Lorentzian transverse field
when periodic continued fractions emerge.

Consider at first, a non-random case.
For a regular alternating chain
\linebreak
$\Omega_1I_1\Omega_2I_2\Omega_3I_3
\Omega_1I_1\Omega_2I_2\Omega_3I_3\ldots$
one generates periodic continued fractions having a period 3,
and after some calculations one gets
\begin{eqnarray}
\rho(\omega)
=\left\{
\begin{array}{ll}
0,
& {\mbox{if}}\;\;\;
\omega<c_6,\;\;\;
c_5<\omega<c_4,\;\;\;
c_3<\omega<c_2,\;\;\;
c_1<\omega,
\\
\frac{1}{3\pi}
\frac{\mid{\cal{X}}(\omega)\mid}
{\sqrt{{\cal{C}}(\omega)}},
& {\mbox{if}}\;\;\;
c_6<\omega<c_5,\;\;\;
c_4<\omega<c_3,\;\;\;
c_2<\omega<c_1;
\end{array}
\right.
\nonumber\\
{\cal{X}}(\omega)
=I_1^2+I_2^2+I_3^2
-\left(\omega-\Omega_1\right)\left(\omega-\Omega_2\right)
-\left(\omega-\Omega_1\right)\left(\omega-\Omega_3\right)
-\left(\omega-\Omega_2\right)\left(\omega-\Omega_3\right),
\nonumber\\
{\cal{C}}(\omega)
=4I_1^2I_2^2I_3^2
\nonumber\\
-\left[
I_1^2\left(\omega-\Omega_3\right)
+I_2^2\left(\omega-\Omega_1\right)
+I_3^2\left(\omega-\Omega_2\right)
-\left(\omega-\Omega_1\right)
\left(\omega-\Omega_2\right)
\left(\omega-\Omega_3\right)
\right]^2
\nonumber\\
=-\prod_{j=1}^6\left(\omega-c_j\right).
\end{eqnarray}
Here $c_1\ge\ldots \ge c_6$
denote six roots of the equation ${\cal{C}}(\omega)=0$
that can be found by solving two cubic equations.
For a regular alternating chain
$\Omega_1I_1\Omega_2I_2\Omega_3I_3\Omega_4I_4
\Omega_1I_1\Omega_2I_2\Omega_3I_3\Omega_4I_4\ldots$
periodic continued fractions having period 4 emerge,
and after some calculations one finds
\begin{eqnarray}
\rho(\omega)
=\left\{
\begin{array}{ll}
0,
& {\mbox{if}}\;\;\;
\omega<d_8,\;\;\;
d_7<\omega<d_6,\;\;\;
d_5<\omega<d_4,\;\;\;
d_3<\omega<d_2,\;\;\;
d_1<\omega,
\\
\frac{1}{4\pi}
\frac{\mid{\cal{W}}(\omega)\mid}
{\sqrt{{\cal{D}}(\omega)}},
& {\mbox{if}}\;\;\;
d_8<\omega<d_7,\;\;\;
d_6<\omega<d_5,\;\;\;
d_4<\omega<d_3,\;\;\;
d_2<\omega<d_1;
\end{array}
\right.
\nonumber\\
{\cal{W}}(\omega)
=I_1^2(2\omega-\Omega_3-\Omega_4)
+I_2^2(2\omega-\Omega_1-\Omega_4)
+I_3^2(2\omega-\Omega_1-\Omega_2)
+I_4^2(2\omega-\Omega_2-\Omega_3)
\nonumber\\
-(\omega-\Omega_1)(\omega-\Omega_2)(\omega-\Omega_3)
-(\omega-\Omega_1)(\omega-\Omega_2)(\omega-\Omega_4)
\nonumber\\
-(\omega-\Omega_1)(\omega-\Omega_3)(\omega-\Omega_4)
-(\omega-\Omega_2)(\omega-\Omega_3)(\omega-\Omega_4),
\nonumber\\
{\cal{D}}(\omega)
=4I_1^2I_2^2I_3^2I_4^2
-\left[
\left(\omega-\Omega_1\right)
\left(\omega-\Omega_2\right)
\left(\omega-\Omega_3\right)
\left(\omega-\Omega_4\right)
\right.
\nonumber\\
\left.
-I_1^2\left(\omega-\Omega_3\right)\left(\omega-\Omega_4\right)
-I_2^2\left(\omega-\Omega_1\right)\left(\omega-\Omega_4\right)
\right.
\nonumber\\
\left.
-I_3^2\left(\omega-\Omega_1\right)\left(\omega-\Omega_2\right)
-I_4^2\left(\omega-\Omega_2\right)\left(\omega-\Omega_3\right)
\right.
\nonumber\\
\left.
+I_1^2I_3^2+I_2^2I_4^2
\right]^2
=-\prod_{j=1}^8\left(\omega-d_j\right).
\end{eqnarray}
Here, $d_1\ge\ldots\ge d_8$ are the eight roots of the equation
${\cal{D}}(\omega)=0$
that can be found by solving two equations of 4th order.
Let us note that all
$c_j$
and $d_j$
are real
since they can be viewed as eigenvalues of symmetric matrices.$^6$

The simplest periodic nonuniform spin-$\frac{1}{2}$ $XX$ chain
in a random Lorentzian transverse field
$\Omega_{01}\Gamma_1I_1
\Omega_{02}\Gamma_2I_2
\Omega_{01}\Gamma_1I_1
\Omega_{02}\Gamma_2I_2\ldots$
requires a calculation of periodic continued fractions with period 2.
The random-averaged density of states for this case becomes
\begin{eqnarray}
\overline{\rho(\omega)}
=
\frac{1}{2\pi}
\frac{\mid{\cal{Y}}(\omega)\mid}
{{\cal{B}}(\omega)};
\nonumber\\
{\cal{Y}}(\omega)
=(\Gamma_1+\Gamma_2)
\sqrt{\frac{{\cal{B}(\omega)}+{\cal{B}}^{\prime}(\omega)}{2}}
-{\mbox{sgn}}{\cal{B}}^{\prime\prime}(\omega)\;
(2\omega-\Omega_{01}-\Omega_{02})
\sqrt{\frac{{\cal{B}(\omega)}-{\cal{B}}^{\prime}(\omega)}{2}},
\nonumber\\
{\cal{B}}(\omega)
=\sqrt{
\left({\cal{B}}^{\prime}(\omega)\right)^2
+\left({\cal{B}}^{\prime\prime}(\omega)\right)^2},
\nonumber\\
{\cal{B}}^{\prime}(\omega)
=
\left[(\omega-\Omega_{01})(\omega-\Omega_{02})
-\Gamma_1\Gamma_2-I_1^2-I_2^2\right]^2
-\left[(\omega-\Omega_{01})\Gamma_2+
(\omega-\Omega_{02})\Gamma_1\right]^2
-4I_1^2I_2^2,
\nonumber\\
{\cal{B}}^{\prime\prime}(\omega)
=2\left[(\omega-\Omega_{01})(\omega-\Omega_{02})
-\Gamma_1\Gamma_2-I_1^2-I_2^2\right]
\left[(\omega-\Omega_{01})\Gamma_2+
(\omega-\Omega_{02})\Gamma_1\right].
\end{eqnarray}
In principal, there are no difficulties in considering more
complicated periodic nonuniform chains apart from the fact that
the calculations
become somewhat cumbersome.

Let us turn to a discussion of the
results obtained for the density of states.
First, note that in the limit of a uniform transverse field
and exchange coupling, Eqs. (7) and (8) reproduce the well-known result
for the uniform spin-$\frac{1}{2}$ $XX$ chain in a transverse field:
$\rho(\omega)=1/\pi\sqrt{4I^2-(\omega-\Omega)^2}$
if $4I^2-(\omega-\Omega)^2>0$ and $\rho(\omega)=0$, otherwise.
This result for the uniform chain also follows from (9) if
$\Omega_{01}=\Omega_{02}=\Omega,$
$\Gamma_1=\Gamma_2=0,$
$I_1=I_2=I.$
The density of states (9) contains the result for uniform
spin-$\frac{1}{2}$ $XX$ chain in a random Lorentzian transverse field$^4$
$\overline{\rho(\omega)}
=\mp(1/\pi){\mbox{Im}}
\left[1/\sqrt{(\omega\pm{\mbox{i}}\Gamma-\Omega_0)^2-4I^2}\right]$
if
$\Omega_{01}=\Omega_{02}=\Omega_0,$
$\Gamma_1=\Gamma_2=\Gamma,$
$I_1=I_2=I.$
In addition, the density of states (9), in the non-random limit
$\Gamma_1=\Gamma_2=0$,
coincides with the density of states (8) with
$\Omega_1=\Omega_3,$
$\Omega_2=\Omega_4,$
$I_1=I_3,$
$I_2=I_4$,  i.e. for a regular alternarting chain
$\Omega_1I_1\Omega_2I_2\Omega_1I_1\Omega_2I_2\ldots$
as should be expected.

In Figs. 1 - 3 we display the density of states,
together with both the dependencies of the transverse magnetization
\begin{eqnarray}
\overline{m_z}=-\frac{1}{2}\int_{-\infty}^{\infty}
{\mbox{d}}E
\overline{\rho(E)}\tanh\frac{E}{2kT}
\end{eqnarray}
on the transverse field at zero temperature,
and the static transverse linear susceptibility
\begin{eqnarray}
\overline{\chi_{zz}}=-\frac{1}{kT}\int_{-\infty}^{\infty}
{\mbox{d}}E
\overline{\rho(E)}\frac{1}{4\cosh^2\frac{E}{2kT}}
\end{eqnarray}
on temperature,
for few particular chains considered.
Initially, let us discuss a non-random case.
The main result of introducing the nonuniformity
is a splitting of the initial magnon band with the edges
$\Omega-2\mid I\mid,$
$\Omega+2\mid I\mid$
into several subbands.
The edges of the subbands are determined by the roots of equations
${\cal{C}}(\omega)=0$,
${\cal{D}}(\omega)=0$.
$\rho(\omega)$ is positive inside the subbands,
tends to infinity inversely proportionally to the square root of
$\omega$
when $\omega$ approaches the subbands edges,
and is equal to zero outside the subbands.
For special values of parameters, the roots of the equations
that determine the subbands edges may become multiple,
the zeros in denominator and numerator
in the expression for the density
of states cancel each other,
and as a result one observes a smaller number of subbands.
In Figs. 1a, 2a we show
$\rho(\omega)$ for some periodic nonuniform chains
having periods 3 and 4
to demonstrate the energy band scheme in the presence of nonuniformity.
Note that the splitting caused by periodic nonuniformity
is not surprising,
since the periodic nonuniform chain can be viewed as
a uniform chain with a crystalline unit cell
containing several sites of the initial lattice.
On the other hand,
one expects several subbands for a crystal having several
atoms per unit cell
(see for example Ref. 19).
We now turn to the density of states given by (9).
First, note that in the non-random case one finds
two magnon subbands the edges of which are given by
\begin{eqnarray}
\left\{b_1,b_2,b_3,b_4\right\}
=
\left\{
\frac{1}{2}
\left[
\Omega_1+\Omega_2\pm
\sqrt{\left(\Omega_1-\Omega_2\right)^2
+4\left(\mid I_1\mid\pm \mid I_2\mid\right)^2}
\right]
\right\}
\end{eqnarray}
(Fig. 3a).
Introduction of the uniform diagonal Lorentzian disorder
$\Gamma_1=\Gamma_2=\Gamma$ results in smearing out of the edges
of subbands (Fig. 3b).
For extremely nonuniform diagonal Lorentzian disorder
$\Gamma_1=0,$
$\Gamma_2\ne 0$ and with a small strength of disorder
only one subband may be mainly smoothed,
but with increasing of the strength of disorder both subbands
become smeared out (Fig. 3c).

The described dependencies of the density of magnon states
on introduction of periodic nonuniformity and disorder
affects the behaviour of thermodynamic quantities of
the spin model.
For example,
the temperature dependence of the specific heat
\linebreak
$\overline{c}
=\int_{-\infty}^{\infty}
{\mbox{d}}E\overline{\rho(E)}
\left(\frac{E}{2kT}\right)^2
/\cosh^2\frac{E}{2kT}$
for a non-random
periodic chain may exhibit a two-peak structure consisting of
low-temperature and high-temperature peaks.
Let us comment in some detail on the magnetic properties of the
spin chains considered.
The splitting of the magnon band into subbands caused by nonuniformity
has interesting consequences for those properties.
Consider for example the transverse magnetization (10).
Since for $T\rightarrow 0$,
$\tanh\frac{E}{2kT}$ tends either to $-1$ if $E<0$,
or to $1$ if $E>0$,
one immediately finds because of the appearance of subbands
that the low-temperature dependence
of $m_z$ against $\Omega$ in the non-random case
must be composed of sharply increasing parts
(when, with increasing $\Omega$, $E=0$ moves
inside each subband
from its top to its bottom)
separated by horizontal parts
(when, with increasing of $\Omega$, $E=0$ moves
inside the gaps). Evidently
a number of plateaus
in the low-temperature dependence of the transverse magnetization
on transverse field
is determined by a number of subbands.
Every cusp in the the dependence
of the transverse magnetization
on $\Omega$
induces a singularity in the dependence
of the static transverse linear susceptibility (11)
on $\Omega.$
One can easily show that the dependence
$\overline{\chi_{zz}}$ against $\Omega$, at $T=0$,
is the same as the dependence
of $\overline{\rho(\omega)}$ against $\Omega-\omega$.
The latter dependence
can be derived from the densities of states depicted in Figs. 1-3.
The value of $\overline{\chi_{zz}}$ at $T=0$
in the temperature dependence
of $\overline{\chi_{zz}}$ at $\Omega=0$ is determined by the value of
$\overline{\rho(0)}$.
Therefore, the nonuniformity and randomness may essentially effect
this value as well as the temperature dependence
of the static transverse linear susceptibility.
In different temperature
regions one finds both an enhancement
(and even a divergence, as in Fig. 3g),
or suppressing of the curves
$-\overline{\chi_{zz}}$ versus $T$.
This can be seen nicely in Figs. 1c, 2c, 3g, 3h, 3i.

To summarize, using continued fractions we have obtained
rigorously the density of magnon states
for periodic nonuniform
spin-$\frac{1}{2}$ $XX$ chain in a random Lorentzian transverse field.
Continued fraction representation of the solution of Eq. (4)
is extremely useful for calculation of the density of magnon states.
The attractive features of this approach can be seen even
for the uniform case. In this case one omits performing twice the Fourier
transformation
while solving Eq. (4) in a standard manner
and evaluates straightforwardly
the desired
$\overline{\rho(\omega)}$.
The advantages of
the continued fraction approach become clear
while treating periodic chains
already with the smallest period of 2. Periodic
nonuniformity leads to a splitting of the magnon band into subbands
that in its turn leads to the appearance of new cusps and singularities
in the low-temperature dependences
of the transverse magnetization and the static transverse linear susceptibility
on transverse field,
respectively.
In the random case the
spectacular changes in these dependences are smoothed.
Periodic nonuniformity and randomness may either enhance or suppress
the temperature dependence of
$-\overline{\chi_{zz}}$.
Changing the degree of periodic nonuniformity
one may to some extent influence the detailed shapes
$\overline{m_z}$
or
$\overline{\chi_{zz}}$
against $\Omega$,
or
$\overline{\chi_{zz}}$ against $T$.
The described approach may be of considerable use for examining
simple models of spin-Peierls instabilities,
especially in the presence of disorder.
Analysis of the thermodynamic properties of
non-random and random Lorentzian
periodic nonuniform
spin-$\frac{1}{2}$ $XX$ chain in a transverse field
and its stability with respect to a lattice distorsion
will be given in a separate paper.

\vspace{5mm}

The author is grateful to O. Zaburannyi
for discussions and J. W. Tucker for
critical reading of the manuscript and useful comments.
He acknowledges
the financial support of
the Research Assistance Foundation (L'viv).
He is also indebted to Mr. Joseph Kocowsky
for continuous financial support.

\clearpage

\begin{figure}
\epsfxsize=160mm
\epsfbox{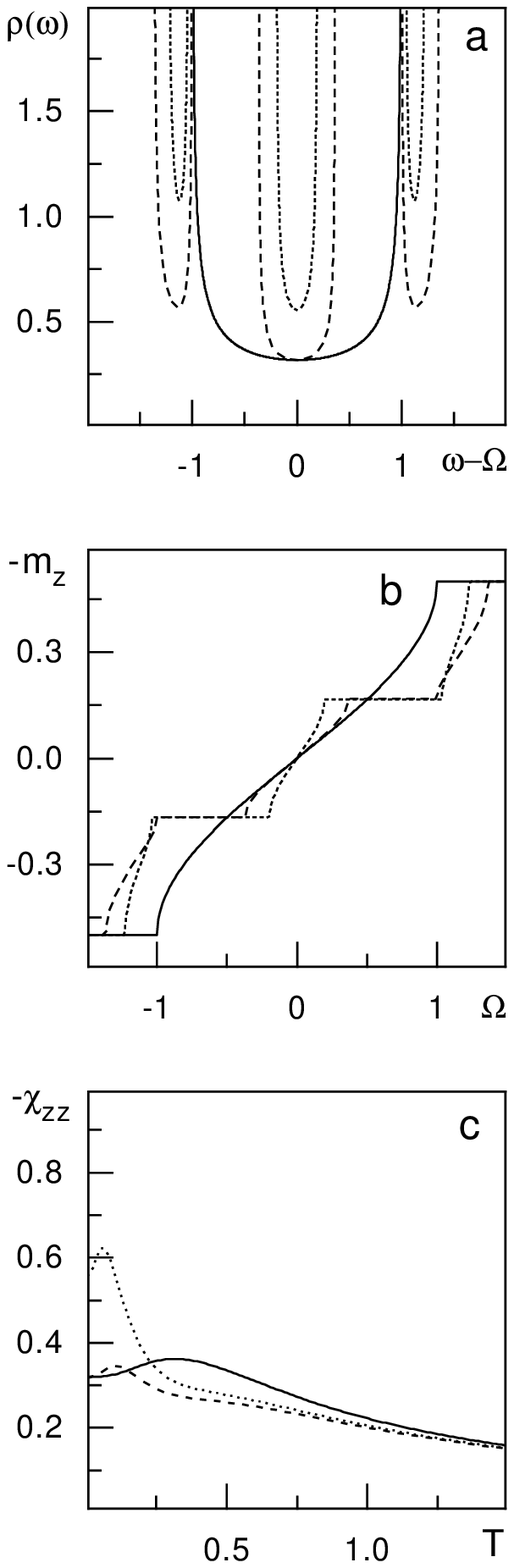}
\vspace{-20mm}
\caption{}
\end{figure}

\clearpage

\begin{figure}
\epsfxsize=160mm
\epsfbox{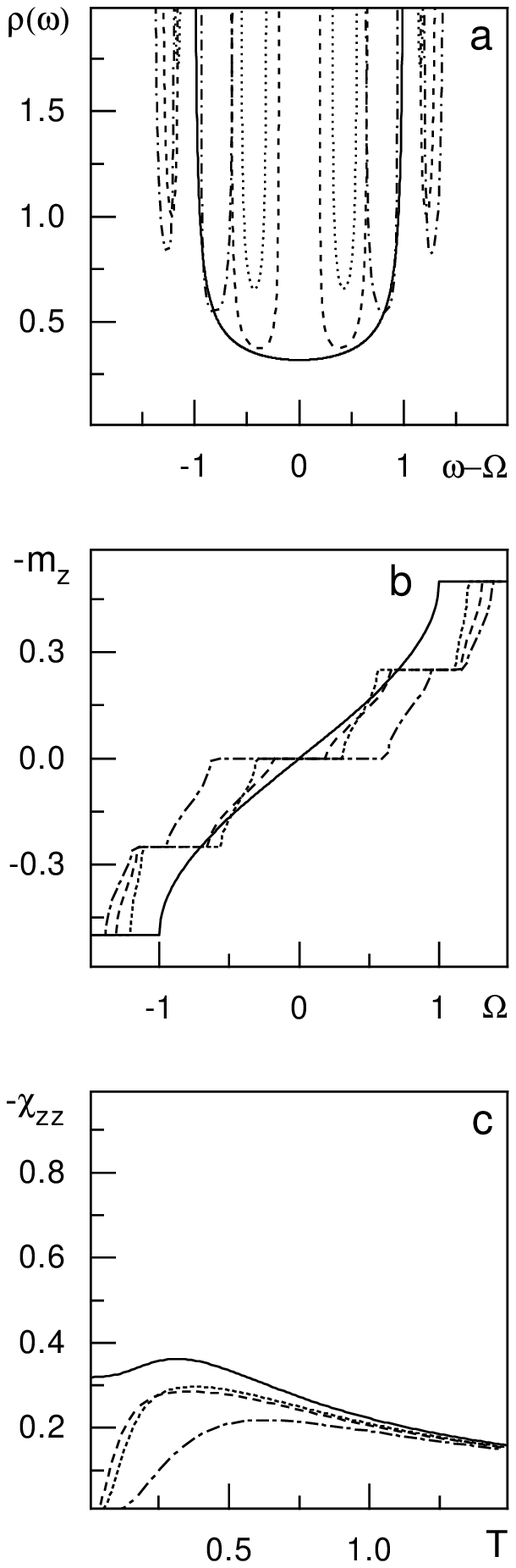}
\vspace{-20mm}
\caption{}
\end{figure}

\clearpage

\begin{figure}
\epsfxsize=160mm
\epsfbox{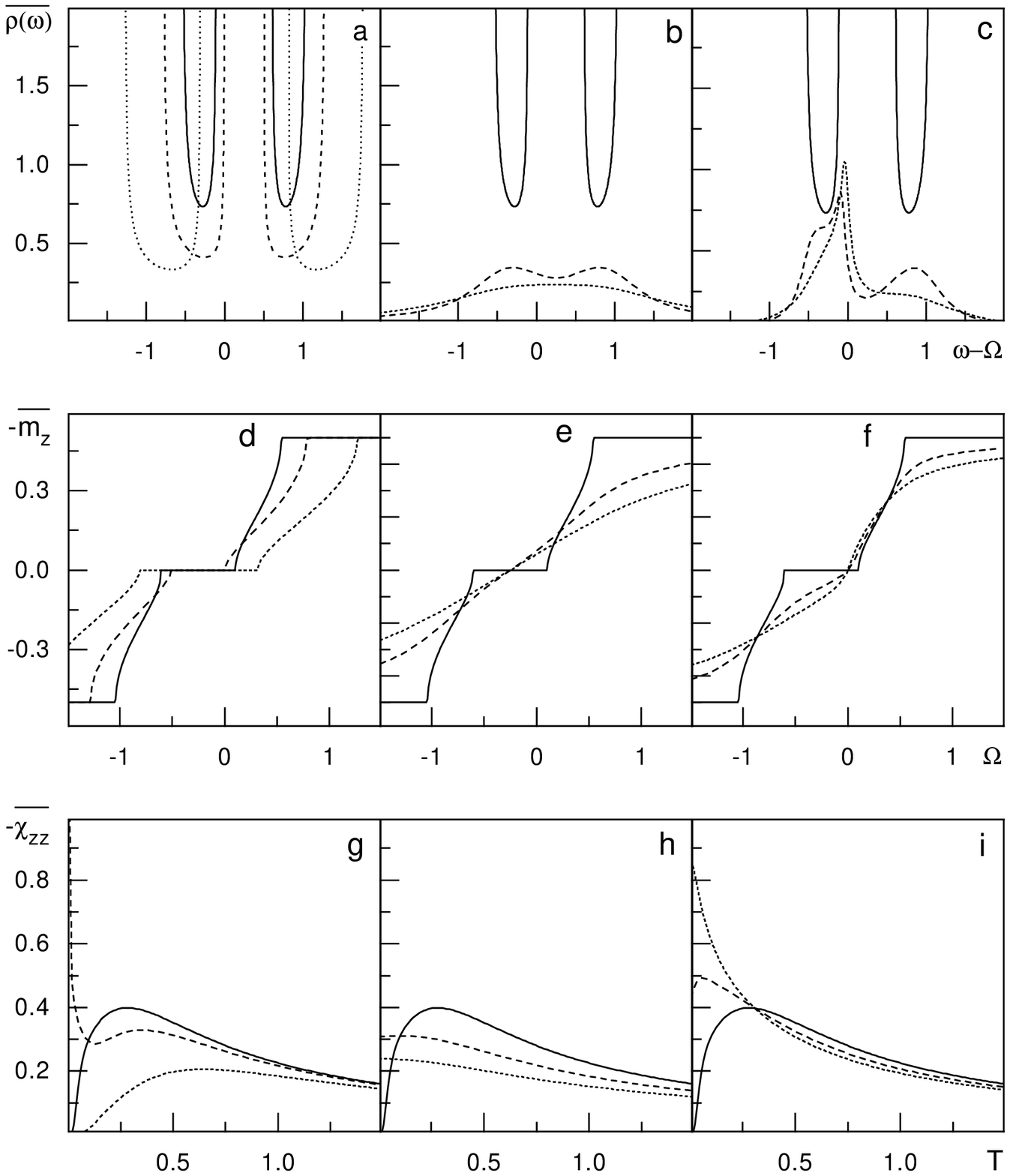}
\vspace{-20mm}
\caption{}
\end{figure}

\clearpage

\noindent
{\bf List of figure captions}

\vspace{1.25cm}

FIG. 1.
Density of states (a),
transverse magnetization
versus transverse field $\Omega$ at $T=0$ (b),
and static transverse linear susceptibility
versus temperature at $\Omega=0$ (c),
for the nonuniform chain
$\Omega I_1\Omega I_2\Omega I_3
\Omega I_1\Omega I_2\Omega I_3\ldots,$
$I_1=0.5.$
$I_2=0.5,$ $I_3=0.5$ (solid
lines);
$I_2=0.5,$ $I_3=1$ (dashed
lines);
$I_2=0.25,$ $I_3=1$ (dotted
lines).

\vspace{1.25cm}

FIG. 2.
The same as in Fig. 1
for the nonuniform chain
$\Omega I_1\Omega I_2\Omega I_3\Omega I_4
\Omega I_1\Omega I_2\Omega I_3\Omega I_4\ldots,$
$I_1=0.5.$
$I_2=0.5,$ $I_3=0.5,$ $I_4=0.5$ (solid
lines);
$I_2=0.5,$ $I_3=0.5,$ $I_4=1$ (dashed
lines);
$I_2=0.5,$ $I_3=0.25,$ $I_4=1$ (dotted
lines);
$I_2=1,$ $I_3=0.25,$ $I_4=1$ (dash-dotted
lines).

\vspace{1.25cm}

FIG. 3.
Density of states (a,b,c),
transverse magnetization
versus transverse field $\Omega$ at $T=0$ (d,e,f),
and static transverse linear susceptibility
versus temperature at $\Omega=0$ (g,h,i),
for the nonuniform random chain
$\Omega_{01}\Gamma_1I_1\Omega_{02}\Gamma_2I_2
\Omega_{01}\Gamma_1I_1\Omega_{02}\Gamma_2I_2\ldots,$
$I_1=0.5,$
$\Omega_{01}=\Omega,$
$\Omega_{01}=0.5+\Omega.$
a, d, g:
non-random case $\Gamma_1=\Gamma_2=0,$
$I_2=0.25$ (solid
lines),
$I_2=0.5$ (dashed
lines),
$I_2=1$ (dotted
lines);
b, e, h:
uniform disorder
$I_2=0.25,$
$\Gamma_1=\Gamma_2=0$ (solid
lines),
$\Gamma_1=\Gamma_2=0.5$ (dashed
lines),
$\Gamma_1=\Gamma_2=1$ (dotted
lines);
c, f, i:
nonuniform disorder
$I_2=0.25,$
$\Gamma_1=0,$
$\Gamma_2=0$ (solid
lines),
$\Gamma_2=0.5$ (dashed
lines),
$\Gamma_2=1$ (dotted
lines).

\end{document}